# Sovereignty in the digital era: the quest for continuous access to dependable technological capabilities[1]

*Roberto Baldoni*

*National Cybersecurity Agency of Italy and Sapienza University of Rome*

*r.baldoni@acn.gov.it*

*Giuseppe Di Luna*

*Sapienza University of Rome*

*diluna@diag.uniroma1.it*

## Abstract

In an era where economies and societies are deeply integrated into cyberspace, achieving a robust level of digital sovereignty has become an essential goal for nations aiming to preserve their security and strategic political autonomy, particularly during turbulent geopolitical times marked by complex global supply chains of critical technologies that ties systemic rivals. Digital sovereignty is a multifaceted, interdisciplinary, and dynamic pursuit that fundamentally relies on a nation's ability to have continuous access to dependable technological capabilities (CTCs) for storing, transferring, and processing domestically produced data. This paper identifies how access continuity or technological dependability could be threatened by several malicious actions from cyberattacks, supply chain tamperings, political or economic actions. By examining different approaches adopted by countries like the United States, China, and the European Union, we highlight different strategies to get access to CTCs depending on their political, economic and institutional nature.

## Introduction

Globalization has left us with a world where autocracies and democracies are interconnected through global supply chains. At the same time, countries like the BRICS (Brazil, Russia, India, China, and South Africa), along with Iran and North Korea, are actively collaborating to challenge the post-World War II world order, which has largely been led by the US and other democratic Western nations, especially following the collapse of the USSR. This complex geopolitical situation has already led to the Ukraine and the Palestinian conflicts.

---

Since the onset of globalization, digital transformation has integrated most of society and industry, including critical sectors, into a vast and intricate cyberspace. It encompasses the entire internet and connected devices, ranging from simple sensors to advanced supercomputers, along with the software supporting both operations and applications, including Artificial Intelligence (AI). It's not merely a static repository of the world's data, but an ever-evolving entity shaped by new technologies, human interaction, digital communication, and automated processes [1].

Today, cyberspace is the primary driver of wealth generation for nations, particularly through data exploitation. As a result, digitalization has shifted from being just another economic sector to a strategic transformative force across all industries. Social media, cloud technology, and AI are profoundly altering societal structures, redefining business models, and reshaping critical national infrastructures. This centrality has made the technological capabilities of cyberspace pivotal in geopolitical dynamics. Paraphrasing what Vladimir Putin said about AI [2], *whoever controls cyberspace will rule the world*. Any threat to cyberspace raises concerns at the governmental level, escalating into national security issues. For example, cyberattacks could disrupt a country's energy grid, block the stock market, while a tampering on the supply chain of a product could put in danger an entire population or a specific group, see the tampering of Hezbollah's pagers by Israeli intelligence in October 2024 [3].

Cyberspace is composed of products, global platforms, and services developed and managed by private companies. Accessing such capabilities provided by the private sector requires a robust and open collaboration between public and private. Such partnerships can become strategic alliances during conflicts, as seen in Ukraine. U.S. tech giants, like Microsoft, Google, Amazon and Starlink, have played a crucial role in protecting Ukrainian government data through cloud services and securing communications via satellite networks. Constructing large-scale cloud or AI infrastructures would be economically unfeasible for public organizations alone, much like in the space sector. This is why even strategic institutions like the U.S. Army partner with major private players for access to cloud services through the *Joint Warfighting Cloud Capability*. Achieving the necessary service standards for military applications would be unsustainable without these partnerships.

Governments are increasingly focused on asserting control over cyberspace, leading to the concept of digital sovereignty. *Digital sovereignty* refers to a government's right to manage its national cyberspace, including the wealth generated within, the safeguarding of national networks, critical digital infrastructures, and the security of data produced by citizens and businesses [1]. A vulnerability of digital sovereignty is a reduction in the *political strategic autonomy of a country,* that is, a nation's ability to steer political decisions without undue influence from external powers (both private and public). Threats to strategic autonomy undermine a nation's security and the well-being of its citizens.

Having *continuous access to state-of-the-art dependable critical technological capabilities* (CTC) is therefore a necessary condition for a country to build solid digital sovereignty. Dependability can be compromised by cyber or physical tampering of the supply chain of a critical technology, by insiders or by using a critical technology manufactured by a strategic geopolitical rival in vital infrastructures. Continuous access can be obstructed by cyberattacks, political decisions blocking the export of critical devices or materials, predatory acquisitions relocating CTCs to other countries, or profit-driven corporate decisions. Additionally, access may be denied based on the personal beliefs of key individuals. This was exemplified in Ukraine, where Elon Musk, CEO of SpaceX (operator of Starlink), declined a request to extend Starlink coverage to support a Ukrainian operation against the Russian fleet in Crimea, citing concerns about escalating tensions and potential nuclear conflict [4].

This paper explores the threats to a country of not having continuous access to trusted CTC in this post-globalization geopolitics. Specifically, we examine how the strength of the domestic industrial sector, market structures, and the security of supply chains impact dependable access to technological capabilities. Our study reveals how the U.S. and China, through different policies reflecting their political systems (democracy vs. autocracy), are pursuing models where CTCs are provided by multiple domestic companies with trusted supply chains. At the same time, the paper highlights Europe's well-known overdependence on non-EU technology providers, a situation exacerbated by fragmented national security and industrial policies. This lack of cohesion makes it difficult to implement common strategies to limit the presence of not dependable CTCs and to set the condition for the creation of large EU technology providers [5], ultimately making the continuous access to dependable CTC difficult undermining the strategic autonomy and geopolitical influence of both the EU and its member states.

## State-of-Art Critical Technological Capabilities

A capability is considered critical when it is essential to a country's economic stability and national security. Such capabilities typically depend on a complex physical infrastructure required for their production or distribution, a skilled workforce with specialized, time-intensive knowledge, and advanced software products. In this article, we focus on the CTCs listed in Table 1, which, according to [1], represent the essential capabilities needed to achieve a minimum level of digital sovereignty today[2].

The market for CTCs has evolved to be dominated by a few global giants, primarily based in the U.S. and China, whose capitalizations are often comparable to the GDP of large countries. These giants possess state-of-the-art capabilities, enabling them to offer top salaries, launch ambitious projects, and acquire the most promising startups. In addition to these global leaders, the CTC market includes numerous regional providers. While smaller in scale, these regional players can still address national CTC needs, though they often need to partner with global providers to deliver advanced products and services.

The semiconductor market has three segments: design, manufacturing (fabrication), and packaging/testing. Design focuses on chip architecture for CPUs, GPUs, and AI processors. Table 1 lists major U.S., EU, and Chinese providers. Leading U.S. players include NVIDIA, Qualcomm, Intel, and AMD (light blue in Table 1), with NVIDIA leading in AI chips. EU firms, like Infineon, NXP, and STMicroelectronics, focus on niche areas (e.g., automotive, sensors, IoT), while Chinese designers supply industrial and consumer chips mainly for the domestic market (light orange in Table 1). The top global player in semiconductor manufacturing is

---

[2] Table 1 includes only lead firms in the CTC supply chain. It does not list key upstream suppliers such as the European company ASML, whose advanced photolithography machines are essential to the chip manufacturing processes of virtually all major semiconductor firms. Similarly, data center infrastructure vendors such as Supermicro, Dell EMC, and HPE, who provide the servers, storage systems, and networking hardware critical to leading firms in the cloud sector, are also excluded.

Taiwan's TSMC, controlling about 62% of the foundry market, with clients all major semiconductor designers relying on its advanced 3nm technology. Other players in Table 1 are leaders in specific sectors or regions (light orange). In chip packaging, Amkor (U.S.) is the main global player with the Taiwanese ASE technology, while China has regional providers, and the EU lacks significant packaging firms.

No relevant companies specializing in physical layer equipment for cellular communication are based in the US which has slowed its progress in 5G development compared to China. To catch up, the U.S. launched the Open RAN initiative to "cloudify" 5G networks through open interfaces and management [10], aiming to attract investment from major U.S.-based cloud providers to strengthen its position in the global market. The global operating system market is dominated by U.S.-based providers, yet many devices, from mobile phones to servers, run on various Linux distributions, an open-source operating system accessible worldwide. Additional gaps include the absence of global EU-based providers of cloud services and complex AI solutions, while the dominant cybersecurity providers are all US-based.

Table 1: CTC lead-firm providers based in USA, China, the EU (light blue means *global providers*, light orange means *regional/subsector-specific providers*)

| Capability | USA | EU | China |
|---|---|---|---|
| semi conductor design | Nvidia, Intel, AMD, Apple, Qualcomm | Infineon, NXP, STMicroelectronics, | HiSlicon, Unisoc |
| semi conductor manufacturing | Intel, Micron, GlobalFoundries | STMicroelectronics, Infineon | SMIC, YMTC |
| semi conductor packaging | Amkor Technology | no relevant player | JCET Group, TFME |
| Telco satellite constellation | Starlink, Viasat | Eutelsat | developing phase with GalaxySpace and Guowang (a national satellite internet project) |
| Telco IP network | Cisco Systems, Juniper Networks | Nokia | Huawei, Zte |
| Telco cellular network | no relevant player | Nokia, Ericsson | Huawei, Zte |
| Operating Systems | Microsoft, Apple, Google | no relevant player | Huawei |
| Cloud | Microsoft, AWS, Alphabet, Oracle | OVH Cloud, T-Systems, Aruba, City Network, Atos | Baidu, Alibaba, Tencent, Huawei, JD Cloud |
| AI | OpenAI, Google DM, Microsoft, AWS, Meta, Tesla | Nokia, SAP, Mistral, Siemens | Baidu, Alibaba, Tencent, Huawei, SenseTime |
| Cyber security | Palo Alto Networks, Fortinet, CrowdStrike, Cisco Security, Microsoft | Bitdefender, ESET, F-Secure, Avast | Qihoo360, Venustech, NSFOCUS, Hillstone Networks, DBAPP |

## Continuous access to dependable CTCs

To remain competitive and secure, companies across all economic sectors need continuous access to dependable CTCs. If access to these technologies is disrupted, or if the CTCs become not anymore dependable, the national economy could suffer severe impacts, potentially leading to collapse. Therefore, *ensuring long-term, dependable access to CTCs requires comprehensive economic, industrial, education strategies*, alongside strong political commitment. Without such measures, a nation risks becoming reliant on private companies or other governments, which compromises its sovereignty and autonomy.

### Threats to dependable CTC

Defining *a dependable CTC involves technological, human, and geopolitical dimensions: it requires evaluating the technological reliability of the product or service, the integrity of the personnel, and the trustworthiness of companies within the supply chain*.

*The Reliability* can be compromised by cyberattacks that exploit software vulnerabilities within any company in the supply chain. For example, in the 2020 SolarWinds breach, attackers infiltrated the software updates of a widely used IT management platform, affecting thousands of organizations worldwide and exposing them to further exploitation. Also if the technology is not state of the art is a reliability issue that can bring to major incidents. Costa Rica, lacking robust security measures for its information systems, suffered a major ransomware attack in 2022 by the Conti group, which halted essential government functions for months [6].

The *availability and integrity of personnel* can also be compromised by insiders, spies, or manipulated employees who facilitate breaches. The sabotage of Hezbollah's beepers illustrates a breakdown in both reliability and integrity: the supply chain's reliability was compromised by substituting the original firmware (a reliability issue), while personnel within one of the supply chain companies enabled physical tampering, embedding a small explosive device intended for Hezbollah operatives. On the personnel availability side, If the specialized workforce of a country falls below a critical threshold, for instance, due to a limited number of STEM graduates or emigration of existing talent, dependability and continuous access may be compromised. This shortage can lead progressively to an increased reliance on foreign CTC providers, ultimately risking a lock-in effect with a consequent restriction of the country's strategic autonomy.

*Trustworthiness*, meanwhile, can be undermined if any entity within the supply chain fails to perform or behave as expected for geopolitical reasons. When companies within a supply chain are based in *like-minded countries*, ensuring *trustworthiness* becomes easier, as companies across the supply chain are more likely to implement aligned standards for security, safety, and maintainability. In such cases, the CTC provider is considered "trusted". Conversely, if the supply chain includes companies from countries with differing or adversarial political agendas, the CTC provider may be deemed "untrusted." It is thus clear that trust is an attribute of the relationship between a provider and a country. For example, Microsoft is a trusted global provider for the U.S. and an untrusted global provider for China.

It is possible, however, to use an untrusted CTC with careful mitigation strategies. For instance, confidentiality risks posed by untrusted telecom equipment can be managed through end-to-end cryptographic solutions or by an untrusted cloud provider through end-user data encryption, privacy preserving computing (such as secret sharing, secure multiparty computation, homomorphic cryptography, ...) and secure hardware enclave. In contrast, availability risks - where an untrusted provider may cease supplying the CTC due to geopolitical pressures - are harder to mitigate. This is why the U.S. banned Huawei and ZTE from national networks in 2019 and it is carrying out an action of cleaning the networks of rural US telecom operators from Chinese devices.

## Threats to continuous access to CTC

Several threats may hamper the continuous access to CTC. Natural disasters have to be considered, for example storms and droughts have caused a restricted chip production in Texas and Taiwan that exacerbated the recent chip shortage.

Political decisions of competing/hostile nations are also a factor to take into account. The U.S. holds a competitive advantage in AI microchips against China and has enacted measures to maintain this edge, including export bans on advanced AI chips, restrictions on skilled personnel transfers, and limits on U.S. investments in Chinese AI. These rules also apply to foreign firms like TSMC, which is barred from supplying advanced chips (7nm and below) to Chinese companies. The U.S. is obstructing China from accessing advanced AI chips.

Additionally, continuous access to technology could be compromised by predatory acquisitions of critical technology assets by systemic rivals. For example, the acquisition of a 40% stake of Epic Games by Chinese Tencent in 2012 raised questions in the US about data privacy and Chinese influence over digital entertainment. Conversely, the acquisition of Qualcomm from Broadcom, a Singapore-based semiconductor company with ties to China, has been blocked by the Trump Administration on the grounds that it threatened the national security of the United States in 2017.

The limited number of global market operators presents another risk to continuous access to CTCs. An incident on such platforms might affect huge areas of the global economy. In July 2024, a widespread disruption occurred when Microsoft Cloud services became inaccessible due to a bugged software update from a CrowdStrike security application, affecting over 8 million computers worldwide. Financial services, airports, and retailers experienced operational disruptions lasting several days. This incident highlights the importance of reducing reliance on a small group of large providers by fostering a more diverse ecosystem of providers and enhancing interoperability to avoid customer lock-in. Such diversity mitigates the impact of similar incidents [7]. The extreme case of lock-in dependency is exemplified by the situation involving Elon Musk, discussed in the Introduction, where Ukraine's strategic autonomy was limited due to dependence on a foreign private individual denying access to a CTC.

We want to remark that CTCs rely on supply chains spanning multiple nations. This acts as a multiplier of both threats described above as any step of the chain has to be secured to ensure a continuous access. Regional providers of CTCs might have shorter supply chains suffering less from this aspect.

## How to mitigate the risks

The foundational step to mitigate the risk of certain CTCs becoming inaccessible is to ensure *diversity* among providers. Therefore, a nation should ensure to have redundancy in the CTC they use. With a diversified provider base, even if a cyberattack, a physical attack, a supply chain attack or accidental incident disrupts one provider, the country can continue to access the CTC. The ideal scenario is for a country to host multiple global providers offering dependable CTCs to ensure continuity of access. When no global provider is domestically based, it is crucial to have at least one regional, state-of-the-art provider with a trusted supply chain. As said, regional providers often need to partner with global providers; in these cases, mechanisms are necessary to ensure that global providers operate in alignment with the country's laws and regulations.

Most global cloud providers are based in the U.S. or China. U.S.-based providers are subject to the CLOUD Act, which allows U.S. authorities to request access to data - including data of foreign nationals or entities - stored in data centers domestically or abroad, based on a judicial order without necessarily notifying the government of the data owner's country [8]. Similarly, following the 2023 revision of China's counterespionage law, Chinese cloud providers must cooperate with Chinese intelligence agencies upon request [9]. To uphold data sovereignty, such data-sharing practices should be either prevented or agreed between countries. If a nation relies solely on foreign CTC providers, ensuring compliance with national cybersecurity or privacy laws becomes significantly more challenging. At minimum, it is essential to diversify with multiple global or regional providers - preferably trusted ones - and to develop advanced expertise within government agencies to set security measures, design advanced projects for the security of sensitive government data and closely monitor provider operations. If such expertise is not available due to skill shortage, the country becomes subordinate to these private companies and, by extension, to the governments where these companies are based, limiting its strategic autonomy.

Achieving diversity and dependability among CTC providers comes at a cost, requiring various strategic long term measures needing firm political will. These measures - whether economic, industrial, or technological - primarily aim to increase the number of providers for each CTC and to secure dependable CTCs. While some actions enhance both accessibility and dependability, others may focus on only one of these aspects.

**Reconfiguring Supply Chains:** Enhancing the dependability of CTCs involves removing untrusted providers from supply chains and replacing them with domestic companies or firms based in like-minded countries. This approach requires robust, long-term economic and industrial policies. For instance, China's drive for technological autonomy, that is all CTCs' supply chains are domestic, reflects its aim to reduce dependence on Western technology. Similarly, the U.S. has implemented policies like the CHIPS and Science Act and the Inflation Reduction Act, which allocate around $500 billions in public incentives to bring essential production to the U.S and to reduce its dependence from China. The EU Chips Act also aims to bolster semiconductor production in Europe, though with considerably fewer resources and slowed down by the EU's more politically and industrial fragmented landscape. Thus, CTC supply chain restructuring for national security is occurring at different speeds in both Western democracies and autocratic nations with the aim also of establishing more CTC national providers.

**Market Expansion:** Expanding the number of market operators can improve both access and dependability of CTCs. Market regulators, such as antitrust agencies, can prevent monopolistic practices, while economic incentives can encourage new providers. However, because competition is global, restricting a major company's domestic operations could weaken its international standing, potentially limiting its ability to support national cybersecurity efforts. Antitrust actions vary by region. In China, regulators have curbed the expansion of domestic tech giants like Alibaba, Tencent, and Didi, aiming to rein in the influence of private capital on social stability. This regulatory crackdown has cost China's tech sector over $1 trillion in market value, equivalent to the economy of the Netherlands. By contrast, EU antitrust efforts have focused on increasing competition by breaking dominant positions, such as requiring Microsoft to unbundle Internet Explorer from Windows Operating System or the recent request to Apple to unbundle the Apple Store App from iOS Operating System allowing concurrency among App Stores. The U.S. has so far taken limited action against its own tech giants, though scrutiny has increased in recent years.

**Regulating acquisitions:** Countries face a delicate balance between attracting foreign investment and protecting national security, economic sovereignty, and technological leadership. Key strategies include setting ownership limits for foreign entities in strategic sectors, conducting thorough due diligence on foreign investors, and retaining veto power over investments that pose security risks. In the U.S., the Committee on Foreign Investment (CFIUS) has the power to scrutinize and block deals between firms (also non-national) if they impact U.S. security interests. CFIUS ordered Beijing Shiji, a Chinese hospitality tech company, to divest its ownership in StayNTouch, a U.S. hotel management software provider. Concerns included potential Chinese access to sensitive data related to U.S. customers and hotel networks. CFIUS prevented a Chinese firm from acquiring Aixtron, a German chip manufacturer (2016) for national security reasons. Similarly, China blocked the acquisition of NXP Semiconductors (a Dutch company) by Qualcomm (an American company) in 2018 because both companies do substantial business in China's technology market.

**Enhancing Software Supply Chain Resilience:** *Software supply chain attacks, which compromise the dependability of CTCs, are a common threat with broad impacts*. To mitigate these risks, it is essential to strengthen the cybersecurity of companies within the supply chain. Key actions include minimizing attack surfaces, enhancing network monitoring for suspicious patterns, strengthening authentication techniques, bolstering employee resilience against social engineering tactics, and increasing cybersecurity awareness among executive boards to address strategic risks effectively from the top down. In addition to this, the transparency and tracking in the software development process must be strengthened across the supply chain, from developers (who create the source code) to suppliers and end-users. One effective tool is a Software Bill of Materials (SBOM), a standardized list of software components, including versions, dependencies, and sources. The U.S. administration recently released draft guidance on SBOMs [10] and has mandated SBOMs for federal contractors and subcontractors [11].

**Creating an Innovation Ecosystem.** An effective innovation ecosystem integrates a highly skilled workforce, sourced from top-ranked tertiary education systems, with funding, infrastructure, and regulatory support to drive growth in technology and business sectors.

Key components include a skilled talent pool, strong venture capital availability, and policies that support research, intellectual property rights, and entrepreneurship. Additionally, government initiatives, research institutions, and industry partnerships are vital to sustaining innovation. Israel serves as a strong example of this model: it combines skilled talent, ample venture funding, and supportive policies, such as R&D incentives, which have made it a global tech leader, particularly in cybersecurity and high-tech startups [13]. Similarly, Silicon Valley in California has built an enduring ecosystem that has continuously fostered groundbreaking technologies. *These ecosystems are essential for creating new companies, and capacities especially in advanced technologies, thereby enhancing the diversity and availability of state-of-the-art providers.*

**Workforce Strategy:** *A skilled workforce is essential for nations to maintain dependable CTCs*. Yet, digital workforce shortages are already noticeable. Cybersecurity Ventures reported 3.5 million unfilled cybersecurity positions globally in 2021, and by 2023, the U.S. had just 65 applicants per 100 digital job openings, a decline from 85 in 2020. This talent shortage poses a growing challenge, though its impact varies by country. Autocratic nations with planned economies, such as China, Russia, and Iran, lead in STEM graduates per capita, with China producing 5 million STEM graduates in 2023. These nations align education with industrial and security needs. In contrast, democracies, which prioritize student choice and institutional autonomy, cannot plan workforce development as tightly. They can however use economic incentive to students and advertisement campaigns to achieve a higher number of STEM graduates.

## Concluding remarks

This paper has argued that continuous access to dependable CTCs is as strategic for a nation as consistent access to energy. We have shown that achieving this requires long-term national economic and industrial strategies backed by firm political determination, rather than isolated technological measures or short-term plans. Democracies face unique challenges in this regard, as frequent political changes, the need for consensus, and public accountability can hinder long-term implementation compared to autocracies, where government stability and centralized decision-making often allow for faster, more cohesive strategic planning.

Additionally, we examined threats to continuous access and dependable CTC provision, identifying key measures to mitigate potential attacks or accidental disruptions. This canvas helps clarify how the U.S. and China are proactively securing continuous CTC access, while highlighting the unique challenges faced by the EU.

The U.S. and China both hold prominent positions in cloud computing, each supported by several major national cloud providers, which strengthens their leadership in AI as well. In the semiconductor sector, as discussed, the U.S. has restricted China's access to advanced microchips, while China has retaliated by imposing export controls on germanium and gallium, two minerals essential for semiconductor manufacturing. Notably, China controls around 62% of germanium and 80% of gallium of the global supply of these minerals, underscoring the strategic impact of these restrictions. China has also implemented strict national security laws that have led some foreign corporations, like IBM, to exit the country, while also banning certain foreign chips and software to enhance dependability in its information systems. When comparing cybersecurity providers, both the U.S. and China have several companies in this space. However, while China's cybersecurity companies primarily serve domestic markets due to national security concerns, U.S. providers often operate on a global scale. The U.S. lacks a national telecom manufacturing giant, which has slowed its progress in 5G development compared to China. To catch up, the U.S. launched the Open RAN initiative to *"cloudify"* 5G networks through open interfaces and management [12], aiming to attract investment from major U.S.-based cloud providers to strengthen its position in the global market.

The EU lags behind for several reasons. Since national security falls under the authority of Member States, there is no cohesive institutional framework for making swift policy decisions on ensuring continuous access to CTCs, unlike in the U.S. or China [1]. This fragmentation also affects industrial policies. Given its authority over the single market, the EU Commission's main tool to address these issues is regulatory, which sometimes leads to excessive market regulation. In addition, the EU's consensus-based decision-making process is complex, and each decision can take years, leaving Europe consistently behind in decision-making - particularly problematic in fast-evolving digital fields - where time is critical. For instance, discussions on the AI Act began in 2019, and the formal legislative path concluded in 2024. Member States then have two additional years to incorporate it into national law, meaning the AI Act will be outdated by 2026, given the rapid advancements in AI with the ChatGPT breakthrough by the end of 2022. In contrast, the U.S. began its AI policy discussions in 2023, following testimony by Sam Altman, CEO of OpenAI, before the U.S. Senate. This quickly led to an AI Executive Order signed by President Biden in November 2023. Historically, the EU has relied primarily on two major global telecom equipment providers, Nokia and Ericsson, while other CTC providers remain regional, limited to individual Member States. This partly stems from strict EU competition rules and inter-Member State rivalries, which hinder the EU's ability to leverage its full market scale to foster industry champions comparable to those in the U.S. or China [5]. Additionally, the EU's open market and fragmented national security policies have allowed both trusted and untrusted foreign providers to capture substantial market shares. This scenario complicates the EU's access to reliable CTCs, ultimately weakening the strategic autonomy and geopolitical influence of both the EU and its Member States in the post-globalization world order.

For nations today, true independence extends beyond physical borders; the digital frontier is vast and complex, demanding continuous access to dependable CTSs and unwavering commitment of governments to trade lengthy bureaucracy for agility, quick fixes for forward-thinking strategies, and insufficient educational priorities for a skilled, future-ready digital workforce to protect their sovereignty. Failing to do so risks forfeiting prosperity and security.